
%
%
%

\magnification = \magstep1


\input epsf
\input rotate

\font\sc=cmcsc10
\font\ninerm = cmr9
\font\ninebf = cmbx9
\font\nineit = cmti9
\font\ninei  = cmmi9
\font\ninesy = cmsy9
\font\ninett = cmtt9

\font\eightrm = cmr8

\font\eightbf = cmbx8
\font\eightsl = cmsl8

\font\eighti  = cmmi8
\font\eightsy = cmsy8
\font\eightex=cmex10 scaled 833

\countdef\refcount=50 \refcount = 0
\countdef\eqcount=100 \eqcount = 0

\def\setrefnum#1{
   \advance\refcount by 1
   \count255 = 50 \advance\count255 by \refcount
   \countdef#1=\count255 #1=\refcount
   }
\def\seteqnum#1{
   \advance\eqcount by 1
   \count255 = 100 \advance\count255 by \eqcount
   \countdef#1=\count255 #1=\eqcount
   }

\newbox\BodyBox
\newbox\CaptionBox
\newbox\RotateBox

\def\makeanemptybody#1{
   \setbox\BodyBox = \vbox to #1 truecm {}
   }

\def\figure#1#2{
   \setbox\CaptionBox=\vtop{
      \textfont0=\eightrm \scriptfont0=\fiverm
      \textfont1=\eighti  \scriptfont1=\fivei
      \textfont2=\eightsy \scriptfont2=\fivesy
      \textfont3=\eightex
      \noindent {\eightbf Figure~{#1}.} \eightrm #2
      }
   \topinsert
      \centerline{\box\BodyBox}
      \smallskip
      \centerline{\box\CaptionBox}
   \endinsert
   }


\def\revtit#1#2#3#4#5{{\it #2} {\bf #3}, #4 (#5).}

\def\author#1{{\sc #1},}

\def\sign{\mathop{\rm sign}\nolimits}
\def\lesssim{\mathrel{\smash{\mathop{<} \limits_{\displaystyle \sim} }}}
\def\gtrsim{\mathrel{\smash{\mathop{>} \limits_{\displaystyle \sim} }}}
\def\sqr#1#2{\vbox{
   \hrule height .#2pt
   \hbox{\vrule width .#2pt height #1pt \kern #1pt
      \vrule width .#2pt}
   \hrule height .#2pt }}
\def\square{\sqr44}
\def\sqrfill#1{\vrule width #1pt height #1pt depth 0pt}
\def\squarefill{\sqrfill4}


\font\titlefont = cmr10 scaled \magstep3
\def\titleline#1{\centerline{\titlefont #1}}

\def\begintitle{\setbox1=\vbox{%
   \null \vskip 0.5 truecm
   \baselineskip = 20pt
   \bgroup}}
\def\endtitle{\egroup \noindent\box1}

\def\name#1{\centerline{\ninerm #1}}
\def\address#1{\centerline{\nineit #1}}
\def\email#1{\centerline{\ninett #1}}

\def\beginauthors{\setbox1=\vbox{%
   \null \vskip 1 truecm
   \baselineskip = 10pt
   \bgroup}}
\def\endauthors{\vskip 1.5 truecm \null \egroup
   \noindent\box1}

\def\beginabstract{ \setbox1 = \vbox{%
   \advance\hsize by -4 truecm
   \baselineskip = 11 pt
   \ninerm
   \textfont0=\ninerm  \scriptfont0=\sevenrm
   \textfont1=\ninei   \scriptfont1=\seveni
   \textfont2=\ninesy  \scriptfont2=\sevensy
   \centerline{\ninebf Abstract}
   \medskip \bgroup \noindent }}
\def\endabstract{ \egroup \centerline{\box1} }

\def\today{$ \rm
   \ifcase\month\or January \or February \or March \or
   April \or May \or June \or July \or August \or
   September \or October \or November \or December \fi \;
   \number\day, \; \number\year $}

\def\preprint#1#2{\line{#1 \hfil INFN-ROM1 #2}}

\def\section#1{\vskip 1 truecm
   \noindent {\bf #1} \medskip}


\setrefnum\ocio
\setrefnum\ritort
\setrefnum\cuku
\setrefnum\dfm
\setrefnum\SZ
\setrefnum\crisanti
\setrefnum\derrida
\setrefnum\opper
\setrefnum\SK
\setrefnum\little
\setrefnum\brunetti
\setrefnum\pablo
\setrefnum\BS
\setrefnum\RM
\setrefnum\omari

\seteqnum\ham
\seteqnum\Jmeas
\seteqnum\one
\seteqnum\two
\seteqnum\nuclei
\seteqnum\Cm
\seteqnum\edfalse
\seteqnum\parity
\seteqnum\edtrue
\seteqnum\Lmod
\seteqnum\three
\seteqnum\LSK
\seteqnum\final
\seteqnum\thermal
\seteqnum\incond
\seteqnum\freezing
\seteqnum\YNT
\seteqnum\relaxation


\nopagenumbers

\preprint{July 22, 1994}{1039}
\line{\hfil cond-mat/9407091}

\begintitle
\titleline{Infinite Volume Relaxation}
\titleline{in the Sherrington-Kirkpatrick Model}
\endtitle

\beginauthors
\name    {Giovanni Ferraro}
\address {Dipartimento di Fisica dell'Universit\`a di Roma ``La Sapienza''}
\address {and INFN -- Sezione di ROMA}
\email   {FERRAROG\%VAXRMA.hepnet@LBL.GOV}
\endauthors

\beginabstract
In a recent work~[\number\opper]
a numerical method has been proposed to simulate off-equilibrium
zero-temperature parallel dynamics for the $SK$ model
without finite size effects. We study the extension of the method to
non-zero temperature and sequential dynamics, and analyze more carefully the
involved computational problems. We find evidence, in the glassy
phase, for
an algebraic relaxation of the energy density to its equilibrium value, at
least at large enough temperatures, and for an algebraic relaxation of the
magnetization to zero at non-zero temperatures,
with an exponent directly proportional to the temperature.
\endabstract

\vfill
\eject
\footline {\hfil \tenrm\folio \hfil}


\line{}
\section{1.~Introduction}

\noindent
In the last years the study of off-equilibrium dynamics for spin glasses
has become of great interest, experimentally, numerically and
theoretically (as a subjective and non exhaustive list of
references see, {\sl e.g.}, [\number\ocio, \number\ritort, \number\cuku]).
The analytical studies are based mostly on the dynamic functional
integral method of de Dominicis [\number\dfm]
($D F M$ in the following) by which, in the infinite
range limit where a mean field solution is exact, the dynamics can be
expressed as a stochastic equation of motion of a {\sl single} spin with
self consistent {\sl nucleus} of evolution and gaussian noise.
In the equilibrium case these equations can be handled imposing
time translation invariance and a suitable way
of recovering partial validity of the fluctuation dissipation theorem,
which is violated in the spin glass phase because of ergodicity breaking.
For such an analysis we recall the fundamental work
of Sompolinsky and Zippelius on the $SK$ model [\number\SZ]
and recent works on the $s S G$ model with
$p$-spin interaction [\number\crisanti].

In spite of the great effort of the last years,
nowadays the analysis of the off-equilibrium
dynamics of mean field models has not been developed to the same extension
as its equilibrium counterpart. From the numerical point of
view the situation is reversed. An initial state for the dynamical
evolution is very difficult to be realized at equilibrium, whereas it is
easy in the off-equilibrium case: the archetypical off-equilibrium
state can be the one in which all the spins of the system point in the same
direction, or else the one in which
the spins point independently in random directions.
The dynamical equations obtained by $D F M$ in the mean field case
are single-spin equations at
thermodynamical limit: this provides, in principle, a powerful tool to
simulate by numerical methods off-equilibrium dynamics without
finite-size effects. However, these equations are not
suitable for a numerical simulation,
as they represent the evolution of continuous spins with continuous time.
A few years ago
in [\number\derrida] it has been developed an approach to the
(parallel) dynamics of mean field models in the case of Ising spins and
discrete times: this approach leads to single-spin equations completely
analogous to the ones obtained by $D F M$ in the continuous case.
Recently, for the $SK$ model at zero temperature,
in [\number\opper] an implementation of these single-spin equations,
written in the form of stochastic finite-difference
equations with independent forcing Gaussian noises,
has been presented.

The aim of this paper is to study more carefully the computational problem
involved in the implementation, introducing a thermal noise at finite
temperature, and discussing the application of the method to sequential
dynamics too.
We prove that the method is reliable for studying
sequential dynamics at every temperature, as far as the relaxation of
physical quantities of the $SK$ model is concerned.
As a first result, in the simplest experimental situation of an evolution
at constant
temperature and zero magnetic field, starting from an initial quench from
the high-temperature paramagnetic state,
we study the relaxation of the energy
density of the model, which was ignored in~[\number\derrida, \number\opper],
together with the magnetization, in the whole range of temperature covering
the glassy phase. We find evidence for
an algebraic relaxation of the energy density to its equilibrium value, at
least at large enough temperatures, and for an algebraic relaxation of the
magnetization to zero at non-zero temperatures,
with an exponent directly proportional to the temperature. This behaviour
for the magnetization is well established experimentally on real spin
glasses (see, for instance, [\number\RM, \number\omari]), but it
has never
been previously tested numerically for mean field models with sufficient
reliability, since so far in all the numerical simulations
it has been difficult to
take finite-size effects into account (see, for instance, [\number\ritort]).

The performance of the
algorithm, in terms of $CPU$-time and memory requirements, is not so
exceptional to qualify it as the {\sl panacea} of the dynamical simulations
for the $SK$ model. However we believe that it is not a remote possibility
to simulate situations directly comparable with the experimental ones
[\number\ocio], which present more delicate
schedules of variation of the physical parameters
({\sl e.g.} temperature and applied magnetic field) and make
observable a phenomenology not confined to a simple relaxation of the
physical quantities.

\medskip

The outline of the paper is the following. In section~2 we discuss the
generalization of the dynamical equations
of~[\number\derrida, \number\opper] to finite temperature, and we justify
their use also for a sequential dynamics,
in section~3 we enter in the details of implementation and precise the
involved computational problems,
in section~4 we present the results of our runs.

\section{2.~Dynamical equations and physical quantities of interest}

\noindent
The Sherrington-Kirkpatrick model for spin glasses ([\number\SK],
$SK$ model in the following) is a model for Ising spins, interacting
{\sl via} a disordered mean field Hamiltonian
$$
H_J [ \sigma ] = - {1\over2} \sum_{i\not=j} \sigma_i J_{i,j} \sigma_j
\eqno(\number\ham)
$$
Here the (symmetric)
couplings $J_{i,j}$ are Gaussian independent random variables, {\sl i.e.}
$$
J_{i,j} = J_{j,i} \quad , \qquad
d\mu(J_{i,j}) = \sqrt{N\over2\pi}
    {\rm e}^{ - N J_{i,j}^2 / 2 } d J_{i,j}
\eqno(\number\Jmeas)
$$
($N$ is the size of the sample; the variance of the $J$'s is $ 1/N $
in order to have finite Hamiltonian density). We denote by
$ \langle \; . \; \rangle $ the thermal average with respect to the
Hamiltonian~(\number\ham) at fixed $J$, and by
$ \overline{\langle \; . \; \rangle}$ the extra average
over the couplings with the measure~(\number\Jmeas).

Our aim is to study a suitable dynamics for this model.
The interesting physical quantities will be
densities of time-varying extensive quantities.
In order to avoid misunderstandings we recall that the couplings $J$ are
constant along the dynamical evolution. Typically we shall consider
the spin-spin correlation function
$$
C (t,t') = \lim_{N\to\infty} { 1 \over N }
\sum_{i=1}^N \overline{\langle \sigma_i(t) \sigma_i(t') \rangle}
$$
and the magnetization
$$
m (t) = \lim_{N\to\infty} { 1 \over N }
\sum_{i=1}^N \overline{\langle \sigma_i(t) \rangle}
$$
If $\sigma_i(t=0) = 1 $ for each site $i$ it is
$$
m(t) = C(t,0)
$$
The energy density is obviously given by
$$
e (t) = \lim_{N\to\infty} { 1 \over N }
\overline{\langle H_J [ \sigma (t) ] \rangle}
$$

We recall now the results of~[\number\derrida, \number\opper].
At {\sl zero temperature} a parallel dynamics for the model is defined
by the equation
$$
\sigma_i (t+1) = \sign [ \sum_{j\not=i} J_{i,j} \sigma_j (t) ] \quad \forall i
\eqno(\number\one)
$$
Averaging over the $J$'s in the limit
$ N\to\infty $, from equation~(\number\one)
in~[\number\derrida, \number\opper]
the following stochastic single-spin equation is derived:
$$
\eqalign{
& \sigma (t+1) = \sign [ h (t) ] \cr
h(t) & = \sum_{s\le t} A_{t,s} \eta (s) + \sum_{s<t} K_{t,s} \sigma (s) \cr
} \eqno(\number\two)
$$
where $\eta$ is a temporary uncorrelated normal noise,
{\sl i.e.} the $\eta(s)$'s at every time $s$ are independent random variables,
each of which is Gaussian distributed with zero mean and variance one.
The {\sl nuclei} $A$ and $K$ in~(\number\two) will be fixed by suitable
self-consistency equations (see~(\number\nuclei) below).

The equation~(\number\two) is equivalent to the~(\number\one) as far as
mean values of density of extensive quantities are concerned, in a meaning
that will be clarified in the following.
In principle equation~(\number\two) can be solved for each realization
of the noise $\eta$, thus giving the time evolution $\sigma(t)$
in function of the noise history $\eta(s)$.
We denote by $\langle \; . \; \rangle_\eta$ the average over the noise
$\eta$. The {\sl nuclei} $A$ and $K$ are fixed
by the self-consistency equations
$$
A_{t,t'} = 0 \; \hbox{for} \; t'>t \quad, \qquad
\sum_{s\le\min(t,t')} A_{t,s} A_{t',s}
= \langle \sigma(t) \sigma(t') \rangle_\eta
\eqno(\number\nuclei.1)
$$
and
$$
K_{t,t'} = 0 \; \hbox{for} \; t'\ge t \quad, \qquad
\sum_{t'\le s < t } K_{t,s} A_{s,t'}
= \langle \sigma(t) \eta(t') \rangle_\eta
\eqno(\number\nuclei.2)
$$
Translated in the formalism of the single-spin
equation~(\number\two) the spin-spin correlation function and the
magnetization become
$$
\eqalign{
& C(t,t') = \langle \sigma(t) \sigma(t') \rangle_\eta \cr
& \quad m(t) = \langle \sigma(t) \rangle_\eta \cr
} \eqno (\number\Cm)
$$
Again, if $\sigma(t=0)=1$, it is
$$
m(t) = C(t,0)
$$
A very high precision plot of the relaxation of $m(t)$ {\sl versus} $t$
until $t=100$ with initial condition $m(0)=1$, is shown in [\number\opper].
The data in~[\number\opper] can well be fitted with an algebraic relaxation
of the type
$$
m(t) \sim m_\infty + A t^{-\alpha}
$$
and the quoted results for $m_\infty$ and $ \alpha $ are
$$
m_\infty = 0.184 \pm 0.002 \qquad \alpha = 0.474 \pm 0.005
$$
The result for $m_\infty$ agrees quite well with what is declared
by~[\number\derrida]
$$
m_\infty = 0.23 \pm 0.02
$$
It is commonly believed that at $ T \not=0 $ this relaxation form becomes
$$
m(t) \sim C t^{-\delta}
$$
with zero remanent magnetization and $ \delta \propto T $. We shall come
back to this point in section~4.

As far as the energy density is concerned, a direct translation in the
formalism of the single-spin equation~(\number\two) would result in
$$
e (t) = - {1\over2} \langle \sigma(t) h(t) \rangle_\eta
\eqno(\number\edfalse)
$$
but this expression is identically zero.
In fact, a peculiar characteristic of the dynamics~(\number\two), already
pointed out in [\number\derrida, \number\opper], is that
$$
\cases{
\langle \sigma(t) \sigma(t') \rangle_\eta = 0 & if $|t-t'|$ is odd \cr
\langle \sigma(t) \eta(t') \rangle_\eta = 0 & if $|t-t'|$ is even \cr
} \eqno(\number\parity)
$$
and it is easy to see that it implies that the form~(\number\edfalse)
for the energy density gives $ e (t) \equiv 0 $.
In [\number\derrida, \number\opper]
this inconsistency problem does not arise,
because the energy density is ignored at all.
We shall see soon that the correct expression is
$$
e (t) = - {1\over2} \langle \sigma(t) h(t-1) \rangle_\eta
      = - {1\over2} \langle \sigma(t-1) h(t) \rangle_\eta
\eqno(\number\edtrue)
$$

\medskip

The source of the problems is the fact that the
parallel dynamics~(\number\one) is not a correct thermal dynamics, even
at $T=0$. In other words it does not drive the system to its
thermal equilibrium, as does its sequential counterpart.
The dynamics~(\number\one) is of practical interest for models of attractor
neural networks, but at a first sight
is not such for the $SK$ model, considered as a thermodynamical model.
Keeping this in mind it is not difficult to accept the fact that
the energy density is identically constant in time, {\sl i.e.} the
dynamics is not dissipative.
We can get rid of this unpleasant situation by considering a slightly
different model, the Little model [\number\little],
for which a parallel dynamics too is a good thermal dynamics.
We shall see soon that
the dynamics defined by~(\number\one), which is a parallel
({\sl i.e.} a bad) dynamics for the $SK$ model, is instead a good thermal
dynamics for the Little model.
In order to define the model
we double the phase space considering, instead
of $N$ Ising spins $\{ \sigma_i \}_i$, $2N$ spins in two sets
$\{ \sigma_i \}_i$ and $\{ \tau_i \}_i$, for $ i = 1 .. N $,
interacting {\sl via} the disordered mean field Hamiltonian
$$
H_J [ \sigma , \tau ] = - \sum_{i\not=j} \sigma_i J_{i,j} \tau_j
\eqno(\number\Lmod)
$$
Again, the (symmetric)
couplings $J$ are Gaussian random variables with the same
measure~(\number\Jmeas). Now, it is easy to see that the (semi)parallel
dynamics
$$
\eqalign{
& \sigma_i (t+1) = \sign [ \sum_{j\not=i} J_{i,j} \tau_j (t) ] \; , \qquad
\tau_i(t+1) = \tau_i(t) \quad \forall i \cr
& \tau_i (t+2) = \sign [ \sum_{j\not=i} \sigma_j (t+1) J_{j,i} ] \; , \;
\sigma_i(t+2) = \sigma_i(t+1) \quad \forall i \cr
} \eqno(\number\three)
$$
is equivalent to a sequential one, thus providing a good thermal dynamics.

In order to recover equation~(\number\one)
we change slightly the notation. Starting from $ t = 0 $,
(\number\three) gives $ \tau_i (t=1) = \tau_i (t=0) $, whereas
the value of $ \sigma_i(t=1) $ is not trivial. Then
$ \sigma_i (t=2) = \sigma_i (t=1) $ and $ \tau_i (t=2) $ is not trivial,
and so on. In general the value of the $ \tau $'s
at odd times and the value of the $ \sigma $'s at even times is trivial, so
these variables are not interesting, as far as the dynamical description of
the system is concerned. Now, we rename {\sl tout court} the remaining
$ \tau $'s (the ones at even times) simply by $ \sigma $. Thus we are left
with an unique set of dynamical variables $ \sigma_i(t) $, and,
rewritten in terms of it,
the equation~(\number\three) is reduced to the previous~(\number\one).

At this point we are left with the following situation: the (formally)
parallel dynamics~(\number\one) is a good thermal dynamics, at zero
temperature, for the Little model. We shall see soon that the introduction
of a finite temperature thermal noise does not present any difficulty.
However, we must firstly justify why we feel us authorized to study the
Little model instead of the SK model. In fact,
the Little model in itself is of little physical interest, but it has
ben shown~[\number\brunetti], both by a replica approach and numerical
simulations, that in the limit $ N \to \infty $ the static properties of
the two models are the same, {\sl i.e.} their thermodynamical behaviours
coincide. In particular, as far as the energy density is concerned, it is
$$
e = \lim_{N\to\infty} { 1 \over N }
\overline{\langle H_J^{(SK)} [ \sigma ] \rangle}
= {1\over2} \lim_{N\to\infty} { 1 \over N }
\overline{\langle H_J^{(Little)} [ \sigma , \tau ] \rangle}
\eqno (\number\LSK)
$$
Now, with our notation, the dependence on time of the Little
hamiltonian~(\number\Lmod) is
$$
H_J^{(Little)} [ \sigma (t) , \tau (t) ]
= - \sum_{i\not=j} \sigma_i(t) J_{i,j} \tau_j(t)
= - \sum_{i\not=j} \sigma_i(t) J_{i,j} \sigma_j(t-1)
$$
Thus it is easy to see that, in the formalism of the single-spin
equation~(\number\two), the energy density becomes
$$
e_{Little} (t) = - \langle \sigma(t) h(t-1) \rangle_\eta
               = - \langle \sigma(t-1) h(t) \rangle_\eta
$$
{}From this expression, and relation~(\number\LSK) between the energies in the
Little and SK models, we recover the previously anticipated
equation~(\number\edtrue) for the correct form of the energy density
(``correct'' in the sense that we can compare correctly the dynamical
results with the static ones for the SK model).

To extend the dynamical equations to finite temperature, it is sufficient
to introduce in the equation~(\number\one) a temperature dependent local
effective field which simulates the thermal agitation.
This field, because of the fact that it does not depend
on the $J$'s, is carried out without modifications during the
derivation of the single-spin equation~(\number\two) starting from
the~(\number\one). The resulting
equation, which generalizes the~(\number\two) of [\number\opper], is
$$
\eqalign{
& \sigma (t+1) = \sign [ h (t) ] \cr
h(t) & = \sum_{s\le t} A_{t,s} \eta (s) + \sum_{s<t} K_{t,s} \sigma (s)
- T \cdot \xi(t) \sigma (t-1) \cr
} \eqno(\number\final)
$$
where the {\sl nuclei} $A$ and $K$ will again be fixed by
the~(\number\nuclei), and
$\eta$ is again a temporary uncorrelated normal noise. The thermal induced
noise $\xi$ is a new temporary uncorrelated noise with law
$$
P(\xi) = \theta(\xi) \cdot 2 {\rm e}^{-2\xi}
\eqno(\number\thermal.1)
$$
which represents a thermal dynamics of {\sl Metropolis} type, or
$$
P(\xi) = {1\over2} [ 1 - \tanh^2 (\xi) ]
\eqno(\number\thermal.2)
$$
which represents a thermal dynamics of {\sl heath-bath} type.
Obviously the presence of this second noise forces
us to slightly change the notation in the above relations~(\number\nuclei),
(\number\Cm) and~(\number\edtrue), now denoting with
$\langle \; . \; \rangle_{\eta,\xi}$ the average over both noises.

Let us stress the fact that in~(\number\final) the thermal field is
proportional to $ \sigma(t-1) $ and not to $ \sigma(t) $. This is a
consequence of the fact that actually we are studying the dynamics of the
Little model and not of the SK model: when we are updating a spin,
deciding if we flip it or not,
its previous value is actually $ \sigma (t-1) $, whereas $ \sigma (t) $
is only a short-hand notation for $ \tau (t) $.
Obviously in the case of the heat-bath dynamics the thermal field can be
reduced simply to $ T \cdot \xi (t) $, as in this case there is no need of
a previous spin to flip: in fact, the distribution law of the
heat-bath noise~(\number\thermal.2) is even in $\xi$. For generality's sake
we maintain the form~(\number\final) for the dynamical equations,
whereas for practical use it is better to pursue in each case simplicity
in place of generality.
In next section we shall see more carefully how to implement
a simulation for the single-spin equation~(\number\final), and
in section~4 we give the results of our runs.

\section{3.~Details of implementation and computational problems}

\noindent
To implement a simulation of the stochastic finite-difference
equation~(\number\final) we follow the main lines of~[\number\opper].
We generate randomly $N_T$ different noise trajectories,
and we define the average over the noise simply by
$$
\langle \sigma(t) \sigma(t') \rangle_{\eta,\xi} \equiv
{1\over N_T} \sum_{i=1}^{N_T} \sigma^{(i)}(t) \sigma^{(i)}(t')
$$
and by analogous expressions for the other interesting quantities. Here
$ \sigma^{(i)}(t) $ denotes the evolution of $\sigma $ in the
$i$-th noise history. Let us stress that
the run-time computation
of certain averages is needed for the evolution itself, since the
{\sl nuclei} $A$ and $K$ are defined in terms of averages over the noise.

We shall be more explicit in a while, let us before anticipate two
comments. Firstly, in the runs which we performed we kept the
temperature fixed, and zero magnetic field.
Changing $T$ and $h$ during the evolution would have been pointless, as
our care was mainly devoted to test the reliability of the method.

Secondly, the choice of the initial conditions
for the evolution is not straightforward.
If in the equation~(\number\final) there was not the
term $ \sigma(t-1) $, as it is for $T=0$ or for finite temperature
heat-bath dynamics, then we would
need only the initial condition at $t=0$, {\sl i.e.}
$ \{ \sigma^{(i)}(t=0) \}_{i=1..N_T} $. In this case, as pointed out
in~[\number\derrida], the choice of the initial condition is to a
great extent arbitrary. In fact, because of the symmetry of the
equation~(\number\final) under simultaneous inversion of $\sigma$ and both
noises $\eta$ and $\xi$, it is not difficult to see that the average of a
product of an even number of fields ($\sigma$ and/or the noises) does not
depend on the initial conditions. The average of a product of an
odd number of fields, on the other hand,
depends on the initial conditions only through a
factor which is simply given by the initial magnetization
$$
m(0) = {1\over N_T} \sum_{i=1}^{N_T} \sigma^{(i)}(t=0)
$$
{}From the physical point of view the initial condition for the evolution
represents a sudden quench from a high temperature paramagnetic
state which is in
equilibrium in a field that sets the value of the magnetization.

In the general case the presence in~(\number\final) of the term
$ \sigma(t-1) $ forces us to impose two distinct initial conditions:
besides $ \{ \sigma^{(i)}(t=0) \}_i $ we need also
$ \{ \sigma^{(i)}(t=-1) \}_i $. At this point it is not clear at all
neither the exact physical meaning of this additional initial
condition, nor how to keep {\sl a priori} under control the dependence of
the evolution on the initial conditions. However, it is reasonable to argue
that the additional initial condition at $ t = -1 $ has no real
significance. We have therefore
fixed the initial conditions in the way we believe
to be the most natural one:
$$
\sigma^{(i)}(t=0) = 1 \qquad \forall i = 1..N_T
\eqno(\number\incond.1)
$$
and
$$
\eqalign{
& \sigma^{(i)}(t=-1) = 1 \qquad \forall i = 1..N_T/2 \cr
& \sigma^{(i)}(t=-1) = -1 \quad \forall i = N_T/2+1 .. N_T \cr
} \eqno(\number\incond.2)
$$
We have not yet performed any systematic test on the dependence
of the evolution on the initial conditions. Let us stress, by the way, that
with the choice~(\number\incond.2) the dynamics preserves the same
property of selecting the parity~(\number\parity) as in the zero
temperature case.

We shall now sketch with some details the algorithm used in the simulation.
Besides the already introduced notation, we denote respectively with
$S$ and $X$ the spin-spin and the spin-noise correlation functions, {\sl i.e.}
$$
S_{t,t'} = \langle \sigma(t) \sigma(t') \rangle_{\eta,\xi}
\qquad
X_{t,t'} = \langle \sigma(t) \eta(t') \rangle_{\eta,\xi}
$$
The main steps are the following:

\item{1.1)} fix the values of: \hfill\break
- type of thermal dynamics (Metropolis or heat-bath) \hfill\break
- temperature $T$ \hfill\break
- number of noise histories $N_T$ \hfill\break
- terminal time point of the evolution $t_{max}$ \hfill\break
- initial seed of the random generator

\item{1.2)} for each history $ i = 1 .. N_T $, fix the initial conditions
for $ \sigma(0) $ and $ \sigma(-1) $ as in~(\number\incond)

\item{1.3)} for each history $ i = 1 .. N_T $, generate the $t=0$ values of
the noises $ \eta(0) $ and $ \xi(0) $ in according, respectively, to a normal
law and to the law~(\number\thermal)

\item{1.4)} set the $ t = 0 $ values of the {\sl nuclei} and of the
correlation functions, {\sl i.e.}
$$
S_{00} = 1 \quad A_{00} = 1 \quad
X_{00} = 0 \quad K_{00} = 0
$$

\item{1.5)} set $ t = 0 $

\item{2)} let $t$ be the current value of the time step; if
$ t \ge t_{max} $ then stop the main execution and go to step 8) (output of
results)

\item{3)} for each history $ i = 1..N_T $, compute $ \sigma(t+1) $
in according to the equation~(\number\final),
by using the {\sl nuclei} $A$ and $K$ already
computed in the previous steps

\item{4)}
compute
$$
\eqalign{
& S_{t+1,s} \equiv
{1\over N_T} \sum_{i=1}^{N_T} \sigma^{(i)}(t+1) \sigma^{(i)}(s)
\quad s < t + 1 \cr
& \Bigl( \; S_{s,t+1} = S_{t+1,s} \; , \; S_{t+1,t+1} = 1 \; \Bigr) \cr
}
$$
and
$$
\eqalign{
& X_{t+1,s} \equiv
{1\over N_T} \sum_{i=1}^{N_T} \sigma^{(i)}(t+1) \eta^{(i)}(s)
\quad s < t + 1 \cr
& \Bigl( \; X_{s,t+1} = 0 \; , \; X_{t+1,t+1} = 0 \; \Bigr) \cr
}
$$
(the correlation functions $ S_{s,s'} $ and $ X_{s,s'} $ for each
$ s, s' \le t $ are already known from the previous steps)

\item{5.1)} compute the {\sl nucleus}
$ A_{t+1,s} $ by inverting its defining equation~(\number\nuclei.1);
this gives the equation
$$
\eqalign{
& A_{t+1,t'} = \sum_{s\le t'} S_{t+1,s} A^{-1}_{t',s} \quad t' \le t \cr
& A_{t+1,t+1} = \sqrt{ 1 - \sum_{t'\le t} \bigl( A_{t+1,t'} \bigr)^2 } \cr
} $$
which is well-posed since we already know
from the previous steps $ A_{s,s'} $, for each $ s, s' \le t $

\item{5.2)} compute the {\sl nucleus}
$ K_{t+1,s} $ by inverting its defining equation~(\number\nuclei.2);
this gives the equation
$$
\eqalign{
& K_{t+1,t'} = \sum_{t'\le s < t+1} X_{t+1,s} A^{-1}_{s,t'} \quad t' \le t \cr
& K_{t+1,t+1} = 0 \cr
} $$
which again is well-posed since we already know
from the previous steps $ A_{s,s'} $, for each $ s, s' \le t $

\item{6)} for each history $ i = 1 .. N_T $, generate the values of
the noises $ \eta(t+1) $ and $ \xi(t+1) $, in according, respectively, to
a normal law and to the law~(\number\thermal)

\item{7)} set $ t = t + 1 $ and go to step 2)

\item{8)} output the results of the run: \hfill\break
- $ S_{t,t'} $ and $ X_{t,t'} $ for $ t, t' \le t_{max} $ \hfill\break
- $ m(t) = S_{t,0} $ \hfill\break
- $ e(t) = - 1/2 \sum_{s<t} A_{t-1,s} X_{t,s} + A_{t,s} X_{t-1,s} $

\smallskip

\noindent
Two technical remarks. Firstly, the random generator for the noises
is the usual machine generator $RAN$ of the VAX-VMS systems. It is useful to
generate two distinct random series for the two noises $\eta$ and $\xi$.
The generator $RAN$ gives an uniform distribution between 0 and 1, but
there is no difficulty to obtain, starting from it, the required
distributions for $ \eta $ and $ \xi $. Secondly,
in steps 5.1 and 5.2 the computation of $ A^{-1} $ is
easily made by the Gauss method thanks to the fact that $A$ is a (lower)
triangular matrix.

As far as the $CPU$-time requiring is concerned, we note that the most
expensive steps are the 3) and the 4), which cost $O(N_T t)$
multiplications and sums, and the 5)'s, which cost $O(t^2)$. Summing over
$t$ until $ t = t_{max} $ we have
$$
CPU \sim c \cdot N_T t_{max}^2 + c' \cdot t_{max}^3
$$
Our runs were performed on a DEC-VAX 6000, running VAX Fortran, where for
$ N_T = 1000 $ and $ t_{max} = 200 $ the required $CPU$-time is about
1 minute.
Concerning the memory occupation, the matrices $S$, $A$ and $X$
require $ O(t_{max}^2) $ memory, whereas $K$ can be stored in a scratch
array of $ O(t_{max}) $, since at time step $t$ we need only
$ \{ K_{t,s} \}_{s<t} $.
The main memory occupation stems however from the complete
$ N_T $ histories of $ \sigma $ and $ \eta $, whereas $ \xi $ is needed
only at current time step $ t $. This gives an occupation of
$ O(N_T t_{max}) $. In conclusion, reducing the memory
occupation as much as possible, we have
$$
Mem/bytes \sim 5 \cdot N_T t_{max} + 3 \cdot t_{max}^2
$$
which gives an occupation of about $1.1$ Mbytes
for $ N_T = 1000 $ and $ t_{max} = 200 $.

Roughly speaking, the $CPU$-time sets a limitation for
$t_{max}$, whereas the memory does it for $N_T$.
If we want to reach a higher value of
$t_{max}$ we only need to wait more time. At a first glance, the
significance of $N_T$ is not so clear. It obviously affects the precision
of the computation, but the error due to finite $N_T$ is not a purely
statistical one, of the order of $ 1 / \sqrt{N_T} $, because average
quantities enter in the evolution itself throughout the {\sl nuclei}
$A$ and $K$.
We cannot thus assume {\sl a priori} that, repeating more times
the same run with a low value of $N_T$, we shall
obtain a higher precision.

It is therefore necessary to understand, at least empirically, the
dependence of the computation results on $N_T$.
This for two reasons: the inherent limitations on a particular
machine force us to keep the value of $N_T$ relatively small
($N_T = 32,000$ at maximum, for our runs); moreover, even if we would
dispose of a bigger machine, we ought to control {\sl a priori} the
reliability of a computation, in order to be confident of its results.
Thanks to the availability of a Cray, in fact, the authors
of~[\number\opper] could keep $ N_T = 1,000,000 $. Such a value appears
sufficiently high, although for no {\sl a priori} reason): we shall come
back to this point in the next section. By the way, though it is not
explicitly declared, probably in~[\number\opper] a slightly different
implementation was used. In fact, $ N_T = 1,000,000 $ would
impose a memory occupation of about 1 Gbyte, which is enormous even for a
Cray. To reduce the memory occupation one could simply avoid to store in
memory the complete $ \sigma $ and $\eta $ histories: one could maintain an
array of $ O(t_{max}) $ with the initial conditions for $ \sigma $ and the
initial seeds of the random generator for $ \eta $, and at every time step
$t$ one computes, or generates, {\sl ex novo}
the $ \sigma $ and $ \eta $ histories until that time, by
using the {\sl nuclei} $A$ and $K$ already computed and stored in memory.
This makes the memory occupation independent on the value of $N_T$, thus
releasing any limitations on it, but increases the $CPU$-time requirement
by a factor $ O(t_{max}) $. On a VAX this would bring to about 25 days
the amount of $CPU$-time spent in a run with $ N_T = 1,000,000 $ and
$ t_{max} = 100 $, but a Cray Y-MP could be even 100 times faster than a
VAX, thus reducing the time to about 6 hours, which is what declared
in~[\number\opper].

We shall now anticipate what we regard as the main source of damage
linked to low values of $N_T$. Let us consider a given history
$ \sigma^{(i)} $. At each time step $t$ there is a probability of spin
flip which, in principle, can depend on the whole history
$ \{ \sigma^{(i)} (s) \}_{s\le t} $. Let us suppose, however, that this
probability is even under global spin inversion. In this case it
is easy to see that
$$
{d m \over dt } = - m(t) p(t)
$$
where $ p(t) $ is the marginal spin flip probability at time $t$.
Now, the probability that none of the $N_T$ histories will flip at time
$t$ is appreciable as soon as $ p(t) N_T \lesssim 1 $. At $T=0$ it is
easy to see that, if at a certain time $\tilde t$ it is
$ \sigma^{(i)} (\tilde t + 1) = \sigma^{(i)} (\tilde t ) $ for each history
$i$, then it is  $ \sigma^{(i)} (t) \equiv \sigma^{(i)} (\tilde t ) $ for
each $ t \ge \tilde t $ and each history $i$, and the system freezes.
At $ T\not=0 $, because of thermal fluctuations, this freezing can never
happen, but we believe that a kind of partial freezing forces the system to
behave anomalously. In next section we shall show such an anomalous
behaviour, though we can not prove that this is due to a mechanism of the
type already described. For the time being, let us note that $ p(t) $ is a
function which decreases to zero as $t$ goes to infinity, at least if
the relaxation of $m(t)$ is slower than exponential (an algebraic law or
a stretched exponential one), as it is commonly believed. Therefore it
becomes $ p(t) N_T \lesssim 1 $ as soon as
$$
t \gtrsim t_{crit} (N_T)
$$
and $t_{crit}$ is an increasing function of $N_T$. It is appealing to
suppose that
$$
m(t) \sim t^{-\alpha}
$$
{\sl i.e.} the relaxation of $ m(t) $ is algebraic to zero, so that
$ p(t) \sim 1/t $ and
$$
t_{crit} (N_T) \sim N_T
\eqno(\number\freezing.1)
$$
Probably there are some additional complications. In fact at $T=0$,
as we anticipated, it is [\number\derrida, \number\opper]
$$
m(t) \sim m_\infty + A t^{-\alpha}
$$
{\sl i.e.} the algebraic relaxation of $ m(t) $ goes to a non-zero value, so
that $ p(t) \sim 1/t^{1+\alpha} $ and
$$
t_{crit} (N_T) \sim N_T^{1 \over 1 + \alpha} \ll N_T
\eqno(\number\freezing.2)
$$
At $ T\not=0 $ it is expected a relaxation to zero, but even if a kind
of freezing mechanism works, it is not clear how to implement it.
We shall see in the next section that, at least qualitatively, the
numerical data give some evidence of the presence of a kind of
``freezing time'' which
grows with $N_T$, though it is difficult to distinguish
between a situation like the one of equation~(\number\freezing.1), where
the freezing time grows quite fast with $N_T$, and
the one of~(\number\freezing.2).

\section{4.~Numerical results}

\noindent
We made two series of runs. The first one was devoted to analyze
the dependence of the results of a run on the value of
$N_T$, in the case $ T = 0.4 $ and thermal dynamics of Metropolis type,
while in the second one we measured the relaxation of the magnetization
and of the energy density in the whole range of temperatures between $0.1$
and $1.5$, for both Metropolis and heath-bath dynamics.
In both series the maximum time reached was $t_{max} = 200 $, and
the external magnetic field was kept equal to zero. We tried also
to measure the relaxation at $ T = 0 $, but for the values of $N_T$ we
used the system freezes before reaching $ t_{max} $.

We discuss firstly the results of the first series. The
choice of the Metropolis dynamics has no particular motivation.
The value of the temperature has been chosen to be
not too high, in order to have relevant effects of the breaking
of the replica symmetry (we remember that with our choice of the
normalization of the coupling~(\number\Jmeas) the glassy critical
temperature is $ T_g = 1 $), and not too low, in order to avoid the very slow
relaxation and the freezing effects of low temperatures.
We made several runs for $N_T$ ranging from $1000$ to $32,000$.
Apart from the systematic error due to $N_T$, which we would investigate,
there is a natural statistical error of order $1/\sqrt{N_T}$, thus
for each value of $N_T$ we made many independent runs, in order to have
comparable statistical errors. More precisely
$$ \vbox { \offinterlineskip \tabskip=3pt
\halign {%
\strut $#$ & $#$ & \hss $#$ & \qquad # & $#$ & \hss $#$ \cr
N_T & = & 1000 & num. of indep. runs & = & 80 \cr
    &   & 2000 &                     &   & 40 \cr
    &   & 4000 &                     &   & 20 \cr
    &   & 8000 &                     &   & 10 \cr
    &   & 16,000 &                     &   & 10 \cr
    &   & 32,000 &                     &   & 10 \cr
}} $$
In figure~1 we plot the energy density $ -2 e(t) $ {\sl versus} $t$
for all the values of $N_T$.
The time scale from $ 0 $ to $ 200 $ was
divided in intervals of size proportionally scaled by a fixed factor
(we have chosen 1.5 to have about ten intervals in the considered
range), then for each run we averaged the data into each interval.
In fact, the original data are strongly dependent at nearby times, so
the form of the plot at small time scales has no significance.
With our choice the reduced points are
equally spaced in logarithmic time scale. This choice for the procedure of
decimation of the data
is due to the fact that we expect an algebraic relaxation,
{\sl i.e.} a linear dependence of $ \ln (e-e_\infty) $ on $ \ln t $, and such
a functional form is invariant under such a transformation, up to an innocent
logarithmic translation in the time scale.
Finally, we made the arithmetic mean of the different runs, and we
kept as error bars for each datum their standard deviation ($ 1 \sigma $).

\setbox\BodyBox = \vbox{
   \epsfysize = \hsize
   \setbox\RotateBox=\hbox{\epsfbox[25 10 550 720] {fig1.ps}}
   \rotr\RotateBox}

\figure{1}{Plot of the relaxation of the energy density $ -2 e(t) $
for Metropolis dynamics
at temperature $ T = 0.4 $ and zero magnetic field. The horizontal dotted
line represents the equilibrium value. Each curve corresponds to a
different value of $N_T$, as it
is shown in the inset, and it has been obtained by averaging the results of
many independent runs, in order to have error bars of comparable magnitude. }

The horizontal dotted line is plotted as a reference and represents
the equilibrium value for the energy density,
computed by numerically solving the replica equations for the $SK$
model~[\number\pablo]. From figure~1
we see that at a certain time the plot of $ -2 e(t) $ starts to grow almost
linearly and rapidly exceeds its equilibrium value.
This anomalous effect is greatly reduced by increasing $N_T$.
It is useless to show a plot
of the magnetization analogous to the one in figure~1, since in this case
the various relaxation curves at different values of $N_T$ do not differ
sufficiently to make the effect of $N_T$ directly observable.
Anyway, if we consider the data at fixed $t$ {\sl versus} $N_T$,
a linear dependence on $ 1/N_T $ clearly appears.
In fact, the data are very well fitted by
$$
Y_{N_T} (t) \simeq Y_\infty (t) + { A(t) \over N_T }
\eqno (\number\YNT)
$$
where $Y$ represents indifferently the energy density $ - 2 e $ or the
magnetization $ m $. The curves $ Y_\infty (t) $, {\sl i.e.} the
extrapolations to $ N_T = \infty $, will be the subject of the next
analysis. Thanks to the very simple dependence~(\number\YNT) on $N_T$, to
obtain a fairly reliable extrapolation of the data $ N_T = \infty $, it is
sufficient to produce two series of measures at two different values of
$N_T$. In the following we shall choose $ N_T = 8000 $ and $ 16,000 $.
The factor $ A(t) $ in~(\number\YNT) depends roughly linearly on $t$, more
precisely:
$$
\eqalign{
A_{energy} (t) & \sim 0.6 \; t \cr
A_{magn.} (t) & \sim 0.2 \, t \cr
} $$
In conclusion, the systematic error due to finite $N_T$, being
$ O(t/N_T) $, is fairly well manageable: at $N_T$ fixed
we can be confident in the results of a run until $t$ remains smaller
than a (not too small) fraction of $N_T$.
It would be very interesting if we understood why the relaxation of the
physical observables presents, for finite $N_T$,
such an anomalous behaviour as in~(\number\YNT),
particularly evident in figure~1 for the energy density.
We believe that the underlying reason is a kind of freezing mechanism
as it is sketched at the end of previous section, but we could
not identify it.

In figure~2 we show the contour lines of the
correlation matrix $ S_{t,t'} $ in the plane $t$-$t'$,
for the different values of
$ N_T $, measured in a randomly chosen run of each set. Obviously at low
values of $N_T$ the plots are very noisy, because of the great statistical
errors. Apart from this noise, the most evident feature of the plots, which
gradually vanishes by increasing $N_T$, is the following.
Let us consider the half plane $ t> t' $, being the figure symmetric.
As soon as $ t$ is greater than
a certain (not well identifiable) critical time,
for each $t'$, $S_{t,t'}$ {\sl versus}
$t$ appears to freeze to a value which is almost constant with $t$.
This behaviour is strongly reminiscent of the above mentioned freezing.

\setbox\BodyBox = \vbox{
\line{\hfil
   \hbox to 0.3\hsize {$ N_T = 1000 $ \hss} \hfil
   \hbox to 0.3\hsize {$ N_T = 2000 $ \hss} \hfil
   \hbox to 0.3\hsize {$ N_T = 4000 $ \hss} \hfil}
\line{\hfil
   \epsfysize = 0.3\hsize
   \setbox\RotateBox=\vbox{\epsfbox[25 10 550 720] {fig2_1.ps}}
   \rotr\RotateBox \hfil
   \epsfysize = 0.3\hsize
   \setbox\RotateBox=\vbox{\epsfbox[25 10 550 720] {fig2_2.ps}}
   \rotr\RotateBox \hfil
   \epsfysize = 0.3\hsize
   \setbox\RotateBox=\vbox{\epsfbox[25 10 550 720] {fig2_3.ps}}
   \rotr\RotateBox \hfil}
\line{\hfil
   \hbox to 0.3\hsize {$ N_T = 8000 $ \hss} \hfil
   \hbox to 0.3\hsize {$ N_T = 16,000 $ \hss} \hfil
   \hbox to 0.3\hsize {$ N_T = 32,000 $ \hss} \hfil}
\line{\hfil
   \epsfysize = 0.3\hsize
   \setbox\RotateBox=\vbox{\epsfbox[25 10 550 720] {fig2_4.ps}}
   \rotr\RotateBox \hfil
   \epsfysize = 0.3\hsize
   \setbox\RotateBox=\vbox{\epsfbox[25 10 550 720] {fig2_5.ps}}
   \rotr\RotateBox \hfil
   \epsfysize = 0.3\hsize
   \setbox\RotateBox=\vbox{\epsfbox[25 10 550 720] {fig2_6.ps}}
   \rotr\RotateBox \hfil}
   }

\figure{2}{Contour lines of the correlation matrix $ S_{t,t'} $
in the plane $t$-$t'$, for Metropolis dynamics at $ T = 0.4 $ and zero
magnetic field. The diagonal of each matrix is $ S_{t,t} = 1 $.
We have plotted ten contour lines equally spaced between 0 and 1,
so that the distance between two successive lines is $ 0.111 $.
Each plot corresponds to a randomly selected run
for the indicated value of $N_T$.
The plots at low $N_T$ are very noisy because of the
great statistical error $ 1/\sqrt{N_T} $. }

\medskip

The main series of measures was devoted to explore systematically the
relaxation curves of the energy density and the magnetization in the whole
range of temperatures between $ T = 0.1 $ and $ T = 1.5 $. We recall that
the glassy transition happens at $ T_g = 1 $. For each temperature value,
and both dynamics (Metropolis or heat-bath) we performed two sets
of independent
runs at two different values of $N_T$, to extrapolate the data
at $N_T = \infty $. More precisely
$$ \vbox { \offinterlineskip \tabskip=3pt
\halign {%
\strut $#$ & $#$ & \hss $#$ & \qquad # & $#$ & \hss $#$ \cr
N_T & = & 8000 & num. of indep. runs & = & 20 \cr
    &   & 16,000 &                   &   & 10 \cr
}} $$
(with the exception of $ T = 0.4 $ with Metropolis dynamics, where we
have already enough data).
Now, we have pre-analyzed the data as follows.
Firstly, for each run, we performed on the data the
previously described procedure of decimation,
to obtain a reduced set of almost-independent time points.
Then, for each value of $N_T$, we averaged the data from the different
runs, and the error bars
were computed by the mean standard deviation ($ 1 \sigma $).
Secondly, we extrapolated to $ N_T = \infty $ assuming a linear dependence
on $ 1 / N_T $ as in~(\number\YNT), that is
we computed, for each time point,
$$
Y_\infty = 2 \cdot Y_{16,000} - Y_{8000}
$$
where the error bars are the obvious ones. For uniformity's sake
we performed such an extrapolation even for $T=0.4$ with Metropolis dynamics,
instead of using the result of the previously described
fitting procedure based on 6 different values of $N_T$.

At this point we are left, for each temperature value and dynamics type,
with a unique set of data giving the time
relaxation of the magnetization and
the energy density of the system. The data are very well fitted by the
following functional forms, for $ T \le T_g = 1 $:
as expected, the magnetization shows an algebraic relaxation to zero
$$
m(t) \sim C t^{-\delta}
\eqno (\number\relaxation.1)
$$
and the energy density shows an
algebraic relaxation to a well-defined non-zero value
$$
e(t) \sim e_\infty + C' t^{-\delta'}
\eqno (\number\relaxation.2)
$$
A few technical remarks. The fits were performed minimizing the (naturally
defined) chi-square function. The one for the magnetization can be easily
linearized and performed by standard methods, the one for the energy density,
being highly non-linear,
requires a package for non-quadratic function minimization and error
analysis, in our case the package MINUIT of
the CERN libraries.
To test the reliability of the fits we impose a cutoff $ t_{min} $ at low
times, we consider the relaxation only for $ t > t_{min} $, and we check
if there is a systematic dependence of the fitted parameters on
$ t_{min} $. The fits of the energy density show no evident dependence. The
ones for the magnetization, for low temperature ($ T \le 0.5 $), show a
systematic decrease of the fitted exponent $ \delta $. This fact is commonly
considered a sign of the presence of non-negligible corrections to the
asymptotic behaviour~(\number\relaxation.1).
Thus we adopt, to take into account
such (unknown) corrections, the Berretti-Sokal procedure, described
in~[\number\BS, sections~4.2 and~5.3]: we fit the data with
the functional form
$$
m(t) = C t^{-\delta} \bigl( 1 + { B \over t } \bigr)
$$
for various {\sl fixed} values of $B$, and we look for the range in $B$ for
which there is no systematic dependence of the fitted parameters $ \delta $
and $C$ on $ t_{min} $. In this range of $B$,
for each fitted parameter we select the maximum and
the minimum values: the best fit for the parameter will be simply the
arithmetic mean of these values, whereas their difference will be
considered as the systematic error due to unconsidered corrections to
the leading behaviour,
or to imperfect knowledge of the form of the corrections
(95\% subjective confidence limit as defined in
[\number\BS, footnote 17]), in addition to the usual statistical
error for the fit, at 95\% confidence level (2$\sigma$).

\setbox\BodyBox = \vbox{
   \epsfysize = 0.7\hsize
   \setbox\RotateBox=\hbox{\epsfbox[25 10 550 720] {fig3.ps}}
   \rotr\RotateBox}

\figure{3}{Plot of the asymptotic value $ - 2 e_\infty $ of the energy
density {\eightsl versus} $T$ for Metropolis dynamics at zero magnetic
field. The data where obtained by a fit with the algebraic relaxation
form~(\number\relaxation.2) for $ T \le T_g $, whereas in the paramagnetic
phase for $ T > T_g $ the energy density relaxes almost instantaneously to
its asymptotic value. The full line is the plot of the energy density for
the $SK$ model at equilibrium~[\number\pablo]; the tail for $ T > T_g $ has
the usual paramagnetic form $ e = - 1 / 2 T $. The error bars from the fit
are negligible with respect to the dimension of the dots. }

Now, the results of the whole analysis can be summarized, for both dynamics,
by three functions of temperature: the exponents $ \delta $ and $ \delta' $
of the algebraic relaxation of the magnetization and the
energy density, and the asymptotic value
$ e_\infty $ of the energy density.
The interesting results are the following.

\setbox\BodyBox = \vbox{
   \epsfysize = 0.7\hsize
   \setbox\RotateBox=\hbox{\epsfbox[25 10 550 720] {fig4.ps}}
   \rotr\RotateBox}

\figure{4}{Plot of the exponent $ \delta' $ of the algebraic relaxation of
the energy density {\eightsl versus} $T$, for both Metropolis~(\squarefill)
and heat-bath~(\square) dynamics at zero magnetic field. The data where
obtained by a fit with the form~(\number\relaxation.2) for $ T \le T_g $,
whereas in the paramagnetic region the relaxation is exponential.
The two series of data hereby displayed are slightly
shifted for mere convenience of representation.
The error bars from the fit are at 95\% confidence level ($2\sigma$). }

\item {-} The energy density relaxes to its equilibrium value, at
least for high enough temperatures ($ T > 0.2 \div 0.3 $).
The plot of $ - 2 e_\infty $ {\sl versus} $T$, for Metropolis dynamics,
is shown in figure~3; the corresponding one for heat-bath dynamics is very
similar. The full line is the plot of the energy density
for the $SK$ model at equilibrium, obtained by an exact numerical solution
of the replica equations [\number\pablo]. Recall that for $ T > T_g $ an
exponential relaxation is expected, instead of the algebraic one of
equation~(\number\relaxation.2), and this can not be observed, since the
procedure of decimation we followed to reduce the data does not leave
invariant an exponential relaxation. However, the asymptotic value is almost
instantaneously reached, so there is no need for a fit. Let us remark that,
in fact, our estimate for the asymptotic value of the relaxation is
slightly {\sl lower} than the equilibrium value (in figure~3
minus two times the energy is plotted,
so the experimental points appear above the
equilibrium curve). This is very probably due to some inaccuracy in the
extrapolation to $ N_T = \infty $.

\setbox\BodyBox = \vbox{
   \hbox to 0.7\hsize {$a$ \hss}
   \epsfysize = 0.7\hsize
   \setbox\RotateBox=\hbox{\epsfbox[25 10 550 720] {fig5_a.ps}}
   \rotr\RotateBox
   \hbox to 0.7\hsize {$b$ \hss}
   \epsfysize = 0.7\hsize
   \setbox\RotateBox=\hbox{\epsfbox[25 10 550 720] {fig5_b.ps}}
   \rotr\RotateBox}

\figure{5}{Plot of the exponent $ \delta $ of the algebraic
relaxation of the magnetization
{\eightsl versus} $T$, for both Metropolis~($a$)
and heat-bath~($b$) dynamics at zero magnetic field. The
data where obtained by a fit with the
form~(\number\relaxation.1) for $ T \le T_g $, whereas in the paramagnetic
region the relaxation is exponential.
The dashed line $ \delta (T) = T $ is drawn as a guide for the eye.
The error bars are the systematic ones
of the Berretti-Sokal procedure for
$ T \le 0.5 $ (95\% subjective confidence limit as defined in
[\number\BS, footnote 17]), whereas for $ T > 0.5 $ they are the usual
statistical ones from the fit, at 95\% confidence level (2$\sigma$). }

\item {-} The exponent $\delta'$ of the relaxation of the energy density is
almost constant with $T$. A plot of $ \delta' $ {\sl versus} $T$, for both
dynamics, is shown in figure~4; let us remark that for Metropolis dynamics
it is  $ \delta' \sim 1 $, slowly increasing with $T$, whereas for
heat-bath dynamics it is $ \delta' \sim 0.7 $, again slowly increasing with
$T$, but with a jump at $ T = T_g = 1 $ where it is $ \delta' \simeq 1$.
This behaviour of the exponent $ \delta' $ is somewhat anomalous, since it
differs from one dynamics to the other, whereas we expect an universal
behaviour. Moreover, though our results agree qualitatively with previous
simulations~[\number\ritort], it would be nice if the exponent of the
relaxation go to zero with $T$.
We shall come back to this point in a while.

\item {-} The exponent $\delta$ of the relaxation of the magnetization is
proportional to $T$. This can be seen from a plot of $ \delta $
{\sl versus} $T$, which is shown, for both dynamics, in figure~5. Moreover,
the data for $ \delta $ are very well compatible with a
dependence on $ T $ of the simple form
$$
\delta(T) = T / T_g
$$
though we have not scrupulously tested this by a fit.
Noticeably, this feature has been observed in
experiments on real spin glasses since a long time
(see, for instance, [\number\RM, \number\omari]).
Strong indications for the validity of
this fact also for the mean-field $SK$ model come from recent
numerical simulations~[\number\ritort], but it has never
been previously tested numerically with sufficient
reliability, since so far in all the numerical simulations it has been
difficult to take finite-size effects into account. Let us stress
that the exact numerical solution of the $SK$ dynamics at equilibrium gives
an exponent $ \nu $ for the algebraic relaxation of the magnetization
which tends to a non-zero value when $T$ goes to zero~[\number\pablo]. This
implies that what is observed cannot be explained by the equilibrium
features of the $SK$ dynamics.

\noindent
As a concluding remark, a few words of caution are necessary. Though from
our data we do not observe a strong evidence for the presence of subleading
terms in the relaxation form of the energy density~(\number\relaxation.2),
nevertheless there are some anomalies in the behaviour of the energy that
could be explained by the underlying presence of such subleading
corrections. In fact, we have already pointed out that at low temperatures
($T<0.3$) our estimate of the asymptotic value of the energy density is
definitely higher than the true equilibrium value. Our data (see figure~3)
show a clear trend to the value $ - 2 e_\infty \simeq 1.4 $ at zero
temperature, in agreement with what was observed in previous
simulations~[\number\ritort]. It is expected that at $ T \not= 0 $ this
value relaxes again very slowly to the true equilibrium value, with an
exponent which goes to zero with $T$. However, due to the fact that this
further relaxation at finite temperature is only a few percent of the main
relaxation at zero temperature, one could argue that at low temperatures
and not sufficiently long times the observed exponent of the relaxation is
an effective exponent, with three consequences: the effective exponent is
not universal, and it is different in Metropolis and heat-bath dynamics;
it is almost constant with $T$, and remains different from zero even at low
temperatures (in~[\number\ritort] it was observed
$\delta' \simeq 0.6 \div 0.8 $ for $T$ ranging from $0.2$ to $0.4$); we
miss the estimate for the asymptotic value, which at low temperatures
remains higher than the true equilibrium value.
All those troubles are not present in the relaxation of the magnetization,
since its zero-temperature asymptotic value is
$ m_\infty \simeq 0.2 $~[\number\derrida, \number\opper], so its further
finite temperature relaxation is much more impressive and much more easy to
observe.


\section{Acknowledgements}

\noindent
I am greatly indebted to Giorgio Parisi
for having originally proposed this work, and
for many stimulating discussions
and helpful suggestions throughout all its development.
All numerical computations were performed at Common Center of
Scientific Calculus of the University of Rome and at Scuola Normale Superiore
in Pisa.


\section{References}

\parskip=5pt

\item{[\number\ocio]} \author {F.Lefloch, J.Hamman, M.Ocio and E.Vincent}
      \revtit {Can aging phenomena discriminate between the droplet model
      and a hierarchical description in spin glasses?}
      {Europhys.Lett.} {18} {647} {1992}
   \hfill\break
      \author {J.Hamman, M.Lederman, M.Ocio, R.Orbach and E.Vincent}
      {\it Physica A} {\bf 185}, {278} (1992) and
      {\it Phys.Rev.B} {\bf 44}, {7403} (1991).

\item{[\number\ritort]} \author {G.Parisi and F.Ritort}
      \revtit {The remanent magnetization in spin-glass models}
      {J.Physique I} {3} {969} {1993}

\item{[\number\cuku]} \author {L.Cugliandolo and J.Kurchan}
      \revtit {Analytical solution of the off-equilibrium dynamics of a
      long range spin glass model} {Phys.Rev.Lett.} {71} {173} {1993}
   \hfill\break
      \author {L.Cugliandolo and J.Kurchan}
      ``On the out of equilibrium relaxation of the SK model''
      to be published in {\it J.Phys.A}
      (archived in {\it cond-mat 9311016}).

\item{[\number\dfm]} \author {C.de Dominicis}
      \revtit {Dynamics as a substitute for replicas in systems with
      quenched random impurities} {Phys.Rev.B} {18} {4913} {1978}

\item{[\number\SZ]} \author {H.Sompolinsky}
      \revtit {Time dependent order parameters in spin glasses}
      {Phys.Rev.Lett.} {47} {935} {1981}
   \hfill\break
      \author {H.Sompolinsky and A.Zippelius}
      \revtit {Relaxational dynamics of the EA model and the mean-field
      theory of spin glasses} {Phys.Rev.B} {25} {6860} {1982}

\item{[\number\crisanti]} \author {A.Crisanti, H.Horner and H.-J.Sommers}
      \revtit {The spherical $p$-spin interaction spin glass model:
      the dynamics} {Z.Phys.B} {92} {257} {1993}

\item{[\number\derrida]} \author {E.Gardner, B.Derrida and P.Mottishaw}
      \revtit {Zero temperature parallel dynamics for infinite range spin
      glasses and neural networks} {J.Physique} {48} {741} {1987}

\item{[\number\opper]} \author {H.Eissfeller and M.Opper}
      \revtit {New method for studying the dynamics of disordered spin
      systems without finite-size effects}
      {Phys.Rev.Lett.} {68} {2094} {1992}

\item{[\number\SK]} \author {D.Sherrington and S.Kirkpatrick}
      \revtit {Solvable model of a spin glass}
      {Phys.Rev.Lett.} {35} {1792} {1975}

\item{[\number\little]} \author {W.A.Little}
      \revtit {???} {Math.Biosci.} {19} {101} {1974}

\item{[\number\brunetti]} \author {R.Brunetti, G.Parisi and F.Ritort}
      \revtit {Asymmetric Little spin glass model}
      {Phys.Rev.B} {46} {5339} {1992}

\item{[\number\pablo]} \author {P.Biscari}
      \revtit{Dynamics of the SK model in the spin-glass phase}
      {J.Phys.A} {23} {3861} {1990}

\item{[\number\BS]} \author {A.Berretti and A.D.Sokal}
      \revtit {New Montecarlo method for the self-avoiding walk}
      {J.Stat.Phys.} {40} {483} {1985}

\item{[\number\RM]} \author {J.Ferr\'e, J.Rajchenbach and H.Maletta}
      \revtit {Faraday rotation measurements of time dependent magnetic
      phenomena in insulating spin glasses}
      {J.Appl.Phys.} {52} {1697} {1981}

\item{[\number\omari]} \author {R.Omari, J.J.Prejean and J.Souletie}
      \revtit {The extra dimension $ W = k T \ln (t/\tau_0) $ of phase space
      below the spin glass transition: an experimental study of the
      relaxation of the magnetization at constant field in
      $ \underline{Cu} M n $} {J.Physique} {45} {1809} {1984}

\bye